
\font\Large=cmr10 scaled\magstep2
\font\ninerm=cmr9 scaled 1000
\magnification 1200
\baselineskip=16pt
\null


\def\twelvepoint{
  \font\twelverm=cmr12   \font\twelvei=cmmi12   
  \font\twelvebf=cmbx12  \font\twelvesy=cmsy10 scaled\magstep 1
  \font\twelveit=cmti12  \font\twelvett=cmtt12  \font\twelvesl=cmsl12
  \font\tenrm=cmr10    \font\teni=cmmi10        
  \font\tensy=cmsy10   \font\tenbf=cmbx10
  \font\ninerm=cmr9    \font\ninei=cmmi9        
  \font\ninesy=cmsy9   \font\ninebf=cmbx9
  \font\eightrm=cmr8   \font\eighti=cmmi8       
  \font\eightsy=cmsy8  \font\eightbf=cmbx8
  \font\fiverm=cmr5    \font\fivei=cmmi5        
  \font\fivesy=cmsy5   \font\fivebf=cmbx5
  \textfont0=\twelverm   \scriptfont0=\eightrm  \scriptscriptfont0=\fiverm
  \textfont1=\twelvei    \scriptfont1=\eighti   \scriptscriptfont1=\fivei
  \textfont2=\twelvesy   \scriptfont2=\eightsy  \scriptscriptfont2=\fivesy
  \textfont3=\tenex      \scriptfont3=\tenex    \scriptscriptfont3=\tenex
  \def\rm{\fam0\twelverm}  \def\mit{\fam1}  \def\cal{\fam2}
  \textfont\itfam=\twelveit  \def\it{\fam\itfam\twelveit}
  \textfont\slfam=\twelvesl  \def\sl{\fam\slfam\twelvesl}
  \textfont\ttfam=\twelvett  \def\tt{\fam\ttfam\twelvett}
  \textfont\bffam=\twelvebf  \scriptfont\bffam=\eightbf
  \scriptscriptfont\bffam=\fivebf  \def\bf{\fam\bffam\twelvebf}
  \let\sc=\ninerm  \let\sb=\ninebf
  \setbox\strutbox=\hbox{\vrule height8.5pt depth3.5pt width0pt}
  \def\doublespace{\baselineskip=24pt   \lineskip=0pt \lineskiplimit=-5pt}
  \def\singlespace{\baselineskip=13.5pt \lineskip=0pt \lineskiplimit=-5pt}
  \def\oneandahalf{\baselineskip=18pt   \lineskip=0pt \lineskiplimit=-5pt}
  \parindent=.5in  \parskip=7pt plus 1pt minus 1pt
  \footline={\ifnum\pageno=1\hfil\else\hss\twelverm-- \folio\ --\hss\fi}
  \def\ltsima{$\;\buildrel<\over\sim\;$} \def\simlt{\lower.5ex\hbox{\ltsima}}
  \def\gtsima{$\;\buildrel>\over\sim\;$} \def\simgt{\lower.5ex\hbox{\gtsima}}
  \hsize=170mm\vsize=230mm\rm\raggedbottom\doublespace
}

\def\tablerule{\noalign{\hrule}}




\def\medtype{\let \rm=\tenrm \let\bf=\tenbf
\let\it=\tenit \let\sl=\tensl \let\mus=\tenmus
\baselineskip=12pt minus 1pt
\rm}
\def\bigtype{\let \rm=\twelverm \let\bf=\twelvebf
\let\it=\twelveit \let\sl=\twelvesl \let\mus=\twelvemus
\baselineskip=14pt minus 1pt
\rm}
\def\smalltype{\let \rm=\eightrm \let\bf=\eightbf
\let\it=\eightit \let\sl=\eightsl \let\mus=\eightmus
\baselineskip=9.5pt minus .75pt
\rm}

\newcount\notenumber

\def\note{\medtype \advance\notenumber by 1
\footnote{$c{\the\notenumber}$}}


\def\co{{\rm C}/\ox}
\def\car{\rm C}

\def\dydzo{$\Delta Y / \Delta \ox$}
\def\esp{\vskip12pt}
\def\h{H}

\def\mbh{$m_{\rm bh}$}
\def\mco{$m_{\rm co}$}
\def\mdot{$M_\odot$}

\def\mt{$M_T$}
\def\ox{\rm O}
\def\ratzo{$(Z-\car-\ox)/ \ox$}
\def\vel{${\rm km\ s}^{-1}$}

\def\z{$Z$}
\def\zo{\ox}
\def\zomu{{\ox-{\Large $\mu$}}}
\def\ref{\par\noindent\hangindent=2pc \hangafter=1 }
\def\'#1{\ifx#1i{\accent"13\i}\else{\accent"13#1}\fi}

\def\ie{\rm i. e.}
\def\eg{\rm e. g.}


\centerline{\Large CHEMICAL EVOLUTION}
\vskip 0.2 cm
\centerline{\Large OF}
\vskip 0.2 cm
\centerline{\Large IRREGULAR AND BLUE COMPACT GALAXIES}
\vskip 0.5 cm

\centerline{LETICIA CARIGI}
\centerline{Centro de Investigaciones de Astronom{\'i}a, Apdo. Postal 264}
\centerline{M\'erida $5101-A$, VENEZUELA}
\centerline{carigi@cida.ve}
\esp
\centerline{and}
\esp
\centerline{PEDRO COL\'IN, MANUEL PEIMBERT and ANTONIO SARMIENTO}
\centerline{Instituto de Astronom{\'i}a, UNAM, Apdo. Postal $70-264$}
\centerline{04510 M\'exico, D. F., MEXICO}
\centerline{colin@astroscu.unam.mx, peimbert@astroscu.unam.mx, ansar@astroscu.
unam.mx}
\esp
\centerline{{\it Received \  \  \  \  \  \  \  \  \  \  \  \  \  \  \  \  \  \
\ 1994}}
\esp
{\bf Abstract. } We discuss the chemical evolution of metal poor galaxies
and conclude that their oxygen deficiency is not due to:
the production of black holes by massive stars or a
varying slope of the Initial Mass Function, IMF, at the high-mass end.
A varying IMF at the low-mass end alone or in combination with: ({\it a})
an outflow of oxygen-rich material, ({\it b})
an outflow of well-mixed material, and ({\it c}) the
presence of dark matter that does not participate in the chemical evolution
process, is needed to explain their oxygen deficiency.
Outflow of material rich in \ox\ helps to account
for the large \dydzo\ values derived from these objects, but it works against
explaining the \ratzo\ and \co\ values.
\esp
\noindent {\it Subject headings}: galaxies:abundances - galaxies:evolution -
galaxies:interstellar matter -
 stars:evolution - stars:luminosity function
\esp
\vfill\eject
\centerline{1. INTRODUCTION}
\esp
By studying the \zomu\ diagram for irregular and blue compact
galaxies, where \zo\ is the oxygen abundance mass
fraction and $\mu$ is the gas mass fraction, it is found that for a given
$\mu$ value, the \zo\ value is smaller than the one predicted by the
closed-box model of chemical evolution with an IMF independent of \z. In a
review of this problem by Peimbert, Col\'in, \&
Sarmiento (1994a), it is argued that the small \zo\ values are not due to:
({\it a}) errors in the \zo\ determination, ({\it b}) inflow, ({\it c})
variations of the slope of the IMF at the high-mass end, or ({\it d}) the
production of black holes by massive stars. Alternatively, these authors argue
that the observed values in the \zomu\ diagram could be due to: ({\it a})
outflow, ({\it b}) variations of the IMF at the low-mass end, ({\it c}) the
presence of dark matter that does not participate in the chemical evolution
process, or ({\it d}) errors in the determination of the total mass of the
galaxies, \mt. We shall make quantitative estimates of some of these effects
based on a comparison of chemical evolution models of galaxies with
observations. In addition to the \zo\ and $\mu$ values, we have three
additional observational restrictions that can be used to constrain
the models: the enhancement in the helium to oxygen mass ratio, \dydzo, the
carbon to oxygen mass ratio, \co, and the heavy elements minus carbon and
oxygen to oxygen mass ratio, \ratzo.
\par
In \S~2 we describe the chemical evolution models of galaxies,
in \S~3 we select ten objects from the best observed ones and obtain
representative values for \zo, $\mu$,
\dydzo, C/O, and \ratzo, values that
should be reproduced by the models.
In \S~4 we compare the observational constraints with the predictions made by
closed-box
models with a varying mass threshold, \mbh\ (hereafter $m$ denotes mass in
solar mass units), for the end of the evolution of massive
stars as black holes (\ie, without enriching the interstellar medium, ISM,
with the heavy elements that are ejected when the massive stars explode in
supernovae events, SNe); these models are computed without the assumption
of the
instant recycling approximation, IRA (\ie, taking into account
the time delay involved in the ISM enrichment),
for two families of IMF's, and variable ages.
In \S~5 we present two types of galactic outflow models and
compare them with the observations.
Finally, the discussion and conclusions are presented in \S~6.
\esp
\centerline{2. CHEMICAL EVOLUTION MODEL}
\esp
\centerline{2.1 Main Features}
\esp
Our chemical evolution model for irregular galaxies takes into account
the dependence of the element yields with the metallicity of the parent galaxy,
it is essentially the model presented
by Carigi (1994) for the solar neighborhood with the following modifications:
a well-mixed or an O-rich galactic wind is added, the infall is excluded, and
the initial gas abundances are chosen as pregalactic ($X_p=0.77$, $Y_p=0.23$,
and $Z_p=0$). Our model considers ages of $t_g =$ 0.1, 1, and 10 Gyr; that is,
models are differentiated from each other in the
sense that they reproduce the representative $\mu $ and \ox\ values at
different times: 0.1, 1, and 10 Gyr. Prantzos et al. (1994) have also presented
chemical evolution models of the Galaxy taking into account the effect of
metallicity dependent yields.
\par
The model by Carigi (1994) is an open one in which the yields,
the lifetimes of stars, the masses of the
remnant stars and the masses of SNe progenitors,
all depend on the metallicity of the progenitor cloud. The IRA is not used
and the evolution of H, $^{4}$He, $^{12}$C, $^{16}$O, $^{56}$Fe, and \z\
is modelled in detail; in particular, the He
content does not increase linearly with metallicity.
The star formation rate, $\psi $, is taken proportional to the gas mass:
$
                \psi = \nu M_{gas}.
$
The stellar yields are non-linearly interpolated from the set of
metallicities presented by Maeder (1992), a procedure based on the results by
Maeder (1991). Prantzos et al. (1994)
also point out that a non-linear interpolation for the yields
improves the agreement of model predictions with observations in the Galaxy.
The main sequence lifetimes have been obtained from Schaller
et al. (1992). The stars are divided into two mass intervals: ({\it a})
low-mass stars (LMS), which are those with initial masses, $m_i$, in the
$0.8 < m_i < 7.5$ range, and ({\it b}) high-mass stars (HMS), those with
initial masses in the $7.5 \leq m_i < 120$ interval. For LMS, the C, O, and
\z\ yields are
taken from Renzini \& Voli ($Z=0.004$ and 0.02, $\alpha$ = 1.5, $\eta$ = 1/3,
1981), and the He yield and remnant masses
from Maeder (including stellar winds, 1992).
For HMS, all yields and remnant masses are taken from Maeder (high-mass
loss rate, 1992).
Three types of SNe are taken into account: Ia, Ib, and II. For SNe Ib,
two possible progenitors are considered: Wolf-Rayet stars and
binary systems; for this type of SNe, the
Fe mass enrichment is 0.15 \mdot, while for SNe II it is taken as 0.075 \mdot\
(Nomoto, Shigeyama, \& Tsujimoto, 1990).
\esp
\centerline{2.2 Stellar Models}
\esp
Since the yields calculated by Maeder (1992, 1993) are extensively used in what
follows, we briefly describe the main features of the models by Maeder and
collaborators; these are:
\vfill\eject
- Stars with $m_i$ from 0.1 to 0.8, do not contribute to the
interstellar gas enrichment and only take gas from the ISM when they are
formed; their main sequence lifetimes are greater than the age of the parent
galaxy.

- Stars with initial masses in the interval $0.8 \le m_i < 1.85$ burn He in
degenerate cores
and end their lives as C-O white dwarfs. Theses stars enrich the ISM mainly
with He, C, and N via stellar winds and planetary nebulae events; their main
sequence lifetimes vary from 1 to 25 Gyr depending on initial mass and
metallicity.

- Stars in the interval $1.85 \le m_i < 7.5$ burn He in non-degenerate cores
and when going through the asymptotic giant branch, enrich the ISM with He,
C, and N via stellar winds until their existence ends in a C-O degenerate core
after the ejection of a planetary nebula, their main sequence lifetimes range
from 0.03 to 1 Gyr; these stars, if in binary systems, may produce SNe I and
contribute to the Fe enrichment, mainly.

- Stars in the range $7.5 \le m_i < m_{wr}$ (lower-limit mass for the
formation of a Wolf-Rayet star) burn elements heavier than He and
mainly eject He, C, N, and O via SNe II; their main sequence lifetimes vary
slightly with the initial chemical content and go from 0.003 to 0.03 Gyr.
Appropriate values for $m_{wr}$ are 25 for a solar metallicity, or 75 for a
twentieth of the solar metallicity.

- Stars with initial masses greater than $m_{wr}$ and low metallicity,
contribute
to the ISM enrichment with He, C, and O via SNe Ib and leave a
massive remnant;
for stars in the same mass range and high metallicities, the strong
contribution to He and C, mainly, is via stellar winds and therefore, their
remnants are less massive. Their lifetimes are less than 0.006 Gyr.
\esp
\centerline{2.3 Initial Mass Functions}
\esp
We have considered two initial mass functions: the first one is
an extension of the IMF for the solar neighborhood derived by Kroupa, Tout \&
Gilmore (1993) to lower masses in order to
take into account substellar objects (KTG IMF, hereafter),
and the second one is the Salpeter (1955) IMF (or S IMF). Both IMFs
are normalized to one; that is,
$$
\int_{m_{l}}^{m_{u}} m \xi(m) dm = 1. \eqno(2.1)
$$
The lower and upper limits of the IMFs are considered to be 0.01 and 120,
respectively.

The KTG IMF is given by:

$$\xi (m) = \cases{ 0.510~m^{-1.3} & if $0.01 \leq m < 0.5$ , \cr
                    0.273~m^{-2.2} & if $0.5 \ \leq m < 1.0$ , \cr
                    0.273~m^{-2.7} & if $1.0 \ \leq m < 120$ , \cr}\eqno(2.2)$$

\noindent where $ \xi(m)dm$ is
the number of stars in the mass interval from $m$ to $m + dm$. For masses in
the range $0.08 \leq
m < 0.5$, KTG find that the exponent for the
95 per cent confidence interval varies from in
$-0.70$ to $-1.85$ and the center is given by $-1.3$.
The KTG IMF has the same slope for $m \geq 1$ as the Scalo (1986) IMF, \ie,
the two IMFs have a different slope only for the $m < 1$ values. Though it has
been ruled out by observations of the solar vicinity a while ago
(Scalo 1986), the Salpeter IMF with a $- 2.35$ value for the slope
from $m$ = 0.01 to 120 will be considered for comparison.
\esp
\centerline{3. OBSERVATIONAL CONSTRAINTS}
\esp
We will generate a series of models with the aim of fitting {\Large
$\mu$}, O, \dydzo, C/O, and \ratzo\ for a `typical' irregular galaxy. Thus,
for any given galaxy, we need to know
the values for these five constraints; however, it is not possible to derive
\dydzo\ for a single galaxy, and a determination using only a pair of
galaxies would have a very poor quality; therefore, we will derive
this value using the average ratio from a group of well-observed galaxies. The
other four constraints will be obtained from average
or representative values of the galaxies in this well-observed group.
\esp
\centerline{3.1   \mt, {\Large $\mu$}, and O}
\esp
In Table 1 we present a group of ten galaxies with well-determined {\Large
$\mu$},
O, and {\it Y} values. The data come from the following
sources: ({\it a}) the {\it Y} values are taken from
Lequeux et al. (1979), Pagel (1987), Torres-Peimbert, Peimbert, \& Fierro
(1989), Skillman (1991), Pagel et al. (1992, hereafter PSTE), Skillman and
Kennicutt (1993), and Skillman et al. (1994); ({\it b})
the \z\ values are derived from the observed {\it N}(O)/{\it N}(H) values under
the assumption that O constitutes 54 \% of the total \z\ value (see below), the
{\it N}(O)/{\it N}(H) ratios are taken from Peimbert \& Torres-Peimbert (1976),
Dufour, Shields, \& Talbot (1981), Kunth \& Sargent (1983), Vigroux, Stasinska,
\& Comte (1987), Skillman, Kennicutt, \& Hodge (1989), Garnett (1990),
Schmidt \& Boller (1993), Skillman and Kennicutt (1993), and Skillman et al.
(1994); ({\it c}) the gaseous mass, $M_{gas}$, the total mass, \mt, and their
ratio, {\Large $\mu$}, are
\vfill\eject
\noindent taken from Lequeux et al. (1979), and
Staveley-Smith, Davies, \& Kinman (1992). The colons after the {\Large $\mu$}
values in the first two lines of Table 1 indicate that the \mt\ values are
very uncertain for these two galaxies.
The values of the total mass shown in Table 1 were obtained
dynamically, therefore they include any possible dark matter material, either
baryonic or non-baryonic. Our models assume that \mt\ is due only to: the
gaseous component, stars, substellar objects, and stellar remnants. The outflow
models start with a higher total mass and reach the $M_T$ value at the end of
the evolution of each model. The last
line in Table 1 contains the average values of the data in the previous rows.

\par
Since the average abundance values are calculated linearly,
the relevance of the \ox-poor objects is much less than when the
average values are calculated using the logarithms of the  corresponding data.
The representative value for \dydzo\ is determined by the whole set
of observations and the role of the first two galaxies in our sample is mainly
to help determine the value for the primordial He.
\par
The adopted values in this paper are: log$ \mu = - 0.53$, and
log(\mt/\mdot) = 9.12. These values are slightly different to the average
values presented in Table 1, nevertheless, a discussion on the effect produced
by varying $\mu$ is given in \S~4 and \S~6. The O
values in Table 1 were derived by: ({\it a}) adopting a uniform electron
temperature over the observed H II region, and ({\it b}) neglecting the
fraction of O atoms tied up in dust grains.
\par
{}From observations of H II regions in the solar neighborhood, it has been
found
that temperature variations along the line of sight increase the \ox/\h\ ratio
by about $0.3$ dex (\eg\ Peimbert, Torres-Peimbert, \& Ruiz, 1992;
Peimbert, Storey, \& Torres-Peimbert, 1993a; Peimbert, Torres-Peimbert, \&
Dufour, 1993b). From the study of a set of metal poor H II galaxies,
Campbell (1988)
finds that the O/H ratios based on temperatures derived from the [O III]
$4363/5007$ ratio, should be increased by factors in the 0.03 to 0.3 dex
range, while from a similar set of objects, McGaugh (1991) finds that the
O/H ratios should be increased
by about 0.2 dex due to temperature variations along the line of sight. These
temperature variations could be due to:
({\it a}) the deposition of mechanical energy inside H II regions by stellar
winds and SNe explosions (\eg\ Peimbert, Sarmiento and Fierro, 1991
and references therein), and
({\it b}) the presence of chemical abundance inhomogeneities.
{}From this discussion and based mainly on the results by McGaugh (1991),
Campbell (1988), and Gonz\'alez-Delgado et al. (1994), we shall assume
that the gaseous O/H ratio should be increased
by 0.16 dex due to the temparature structure inside H II regions.
\par
Dufour et al. (1994) have found that the ratio Si/O in irregular galaxies is
from 1.5 to 3
times smaller than the solar value, indicating that an important fraction of
the Si atoms is in the form of dust grains; this result implies that about
10 \% of the O atoms are trapped by Si, Mg and Fe in dust grains.
Consequently, the O abundance should be increased by 0.04 dex. These two
corrections increase the O average value in Table 1 from $1.63 \times
10^{-3}$ to $2.58 \times 10^{-3}$.
\esp
\centerline{3.2   \dydzo}
\esp
{}From a linear regression of the sample, it is found that \dydzo\ $= 7.10
\pm 1.62$.
$\Delta $O should be increased by 0.2 dex to take into account the
temperature structure of the H II regions and the fraction of O embedded in
dust; alternatively, $\Delta Y$ is not affected by these two effects since the
temperature structure diminishes the $Y$ determinations but does not affect
$\Delta Y$ up to a very good approximation, and no He is expected to be in
dust grains. Hence, we obtain for our sample that \dydzo\ $= 4.48 \pm 1.02$,
and in the following we shall adopt log(\dydzo) $= 0.65 \pm 0.11$ as the
representative value to be fitted by our models.
\par
We shall compare our \dydzo\ determinations with those derived by other
authors; in particular with PSTE, one of the most important data sets found
in the literature. From these data it is found that \dydzo\ $= 10.2 \pm 3.5$,
and combining them with the sample by Isotov, Thuan \& Lipovetsky (1994), it
is found that \dydzo\ $= 9.7 \pm 2.8$. These values have been derived under the
assumption of a uniform temperature inside the H II regions and without
considering the amount of O embedded in dust grains, therefore in good
agreement with the values derived from our sample.
\par
{}From a comparison of $Y_p$ with the O and $Y$ abundances of the solar
vicinity,
Peimbert, Sarmiento \& Col{\'i}n (1994b) have determined a value of \dydzo\
$= 5.22 \pm 1.1$, which is in excellent agreement with the value adopted by
us for irregular galaxies.
\esp
\centerline{3.3   \co}
\esp
The \co\ ratio varies with O from a low value of about $- 1.1$ dex for
O-poor objects, to a high value of about $- 0.3$ dex for the solar vicinity
(Garnett et al., 1994; Peimbert, 1993 and references therein). For the \co\
constraint we shall consider well observed objects with O values similar to
those of our `typical' irregular galaxy; these are: 30 Doradus in the LMC and
NGC 2363 in NGC 2366. For 30 Doradus and NGC 2363, Garnett et al. obtain
$- 0.60 \pm 0.26$ dex and $- 0.75 \pm 0.15$ dex, respectively,
while Peimbert, Pe\~na \& Torres-Peimbert (1986) obtain $- 0.65 \pm 0.15$
dex for NGC 2363. These results are not affected by the temperature
structure since the lines used to derive the C and O
abundances have similar excitation energies, and to a very good approximation
the temperature structure cancels out. These three determinations correspond to
the gaseous component and do not include the O trapped in dust grains. From
these determinations and assuming that the fraction of O trapped in dust
grains amounts to 0.04 dex, we shall adopt for the \co\ ratio the value
of $- 0.70 \pm 0.15$ dex.

We are assuming that the fraction of C embedded in
dust grains is negligible. For the MNR/Draine-Lee grains (Mathis, Rumpl \&
Nordsieck, 1977; Draine \& Lee, 1984), made by a combination of silicate and
graphite grains, the strenghts of the 9.7 $\mu$m and 18 $\mu$m absorption
features imply that practically all the Si atoms are in the Si$-$O stretch
and consequently, that the amount of SiC is negligible (Mathis, 1994: private
communication). On the other hand, the 2175 \AA\ bump, narrower in the H II
regions
than in the general ISM, is probably due to the removal of dust coatings by
the higher radiation field and higher grain temperatures (Mathis, 1994), and
the high C/H values in the Orion nebula and M17 (Peimbert, Torres-Peimbert
\& Ruiz, 1992; Peimbert, 1993) also indicate that not much C is trapped by
dust inside H II regions. In any case, a change in the adopted C/O constraint
by 0.1 dex does not modify considerably our main conclusions (see Figures 2
and 5).
\esp
\centerline{3.4   \ratzo}
\esp
To determine the \ratzo\ value, we shall consider again the best
observed irregular galaxies with O values similar to those of our `typical'
irregular galaxy; these objects are 30 Doradus and NGC 2363. For 30 Doradus
we shall take the \co\ ratio from Garnett et al. (1994) and the N, O, Ne, S,
and Ar values presented in Torres-Peimbert et al. (1989), increasing the O
value by 0.04 dex to include the O trapped in dust grains. We shall
assume that the mass ratio of all the other unobserved heavy elements to O
is the same as in the sun: 36 \% (Grevesse \& Anders, 1989; Grevesse et al.,
1990; Bi\'emont et al., 1991; Holweger et al., 1991; and Hannaford et al.,
1992). From these values we obtain \ratzo\ = 0.645 and O/\z\ = 0.534
for 30 Doradus. For NGC 2363 we shall take an average of the C/O ratios by
Garnett et al. (1994) and Peimbert et al. (1986), while for N, O, Ne, S, and
Ar we shall take the values from Peimbert et al. With the same assumptions
as those adopted for 30 Doradus, we obtain that \ratzo\ = 0.660 and
O/\z\ = 0.543; these values can be compared to the solar ones that are 0.726
and 0.488, respectively, the differences are due to the smaller C/O and N/O
values present in the irregular galaxies relative to the solar ones. From
30 Doradus and NGC 2363 we have adopted the value log[\ratzo] = $- 0.18
\pm 0.06$.
\esp
\centerline{4. CLOSED BOX MODELS}
\esp
We have computed a series of closed box models --with and without the
contribution of SNe I, for various IMFs, ages, and CO-core masses at the end of
the C-burning phase, \mco-- in order to
fit the adopted log$ \mu $ value of our sample and two possible values of \ox:
$1.62 \times 10^{-3}$, and $2.58 \times 10^{-3}$. The variation of the CO
core
mass was used in order to study the suggestion by Maeder (1992, 1993), in the
sense that black holes could play an important role in the chemical evolution
of metal poor galaxies; the models assumed that the entire star becomes a black
hole if the mass of the CO-core is larger than \mco.
\par
In Figures 1, 2, and 3, a dot represents a model with an age of 10 Gyr, the KTG
IMF, and \mco\ $= 60$ (that is, without the presence of black holes), that fits
the adopted values: log$ \mu = - 0.53$ and \ox\ $= 2.58 \times 10^{-3}$.
This model considers
the contributions of SNe I of binary origin under the assumption that 5.7
\%  and 2.4 \% of the binary systems become SNe Ia and Ib, respectively;
these are the fractions needed to adjust the current SNe rates  in the solar
vicinity (see Carigi, 1994). The SNe I contributions to \ox\ and \z\ are only
2.3 \%  and 10.1 \% respectively; moreover, the SNe I rates of irregular
galaxies are not well known. For the 0.1 and 1 Gyr series of models, the
contributions of
SNe I to the chemical evolution of galaxies are completely negligible. Based on
the previous discussion, the SNe I have not been taken into account for all the
other models in this paper.
\par
In Figures 1, 2, and 3, we also present three series of models for the KTG IMF
with different ages and one series of models for the S IMF with 10 Gyr. In
Figure 3, the KTG curves for the 1 and 10 Gyr series are almost identical
because the contributions by stars in the $0.9 < m_i < 1.85$ range to the
\ratzo\ ratio are almost negligible. In Figures 1 and 2 the increase of \dydzo\
and \co\ with $t_g$ is due to the C and He enrichment of the ISM by LMS.
The differences presented in Figures 1, 2, and 3 between
the KTG-10 Gyr and the S series are mainly due to the larger number of stars in
the $0.9 < m_i < 7.5$ interval in the KTG IMF compared to that in the S IMF.
\par
Table 2 summarizes the results obtained by eight models
with an age of 10 Gyr, a KTG IMF, and an
initial mass gas of $1.33 \times 10^9 M_\odot$. Seven of these models fit
log$ \mu = - 0.53$ and \ox\ = $2.58 \times 10^{-3}$ (see Table 1),
while the eighth model fits the adopted log$ \mu$ value and
\ox\ = $1.62 \times 10^{-3}$; the changes in the element ratios between the
first and last rows in Table 2 are due to the dependence of the element yields
on the initial \z. The first column
shows \mco\ values (related to \mbh\ according to Maeder, 1992). The two
values present in the second column indicate the lowest stellar mass
for the formation of a black hole when $Z = 0.001$, and to the same limit
when each one
of the models reaches its maximum metallicity. The third column shows the
factor, $r$, by which the number of stars that contribute to the ISM enrichment
has to be reduced in order to fit the adopted log$ \mu$ and \ox\ values.
The remaining columns give the values of
the He, C, and $Z$ abundances by mass, and some combinations of them.
\par
In Table 3 we present the mass budget of the ISM from the different processes
considered in four models with \mco\ $= 60$ (\ie, no black holes present) and
the KTG IMF. The different columns stand for: 1) age of the model, 2) chemical
species, 3) mass lost by star formation (SF), 4) mass ejected by low-mass
stars (LMS), 5) mass ejected by the winds of high-mass stars (HMS$_W$), 6)
mass ejected by SNe explosions (HMS$_{\rm SN}$), and 7) the ISM abundance by
mass. In the contributions by the different stellar ejecta, both,
the material already present and the material synthesized within the stars,
are included.
\par
{}From Table 2 and Figures 1, 2, and 3 it follows that there is not a unique
\mco\ value that fits the three observational restrictions for the 10 Gyr
models. There are two extreme solutions that can be adopted: ({\it a}) we can
fit the \dydzo\ value with \mco\ $= 16.4$ that corresponds to an $r = 1.97$
but without satisfying the \co\ and \ratzo\ restrictions, or ({\it b}) we
can fit \co\ and \ratzo\ with \mco\ $= 60$ and $r = 3.39$ but without
satisfying the \dydzo\ restriction. Note that the S IMF series and the KTG
IMF 0.1 and
1 Gyr series predict even smaller \dydzo\ values than the KTG-10 Gyr series.
Since star formation in the LMC and the SMC has been going on for about 10 Gyr,
we consider the 10 Gyr series more adequate.
\par
Carigi (1994) and Peimbert et al. (1994) have found that the production of
black holes by massive stars does not play an
important role in the chemical evolution of the ISM in the solar vicinity. This
result combined with the \co\ and \ratzo\ constraints, has led us to consider
that the ({\it b}) possibility of the previous paragraph is more likely.
The SNe I
contributions increase the predicted \ratzo\ and \co\ ratios by about 20 \%
and 2 \%, respectively, making the presence of black holes more
unlikely (see Figures 2, and 3).
\par
To reduce the \ox\ production, we have to adopt an IMF with a smaller fraction
of massive stars relative to all the other stars. This can be done by
either steepening the slope of the high-mass end of the IMF, or by increasing
the amount of objects in the $0.01 < m < 1$ interval relative to those in the
$1 < m < 120$ interval.
\par
There are four arguments against changes in the slope at the high-mass end of
the IMF with
metallicity: ({\it a}) the study of the ionization degree of many
extragalactic H II regions in irregular and blue compact galaxies with
different \ox\ content (McGaugh, 1991); ({\it b}) the observed continuum
in the 1100--7500 \AA  range of starburst regions with different \ox\
content in M101 (Rosa \& Benvenuti, 1994); ({\it c}) direct determinations
of the high-mass end slope in the SMC and the LMC (Massey et al., 1989a, b;
Parker et al., 1992; Parker \& Garmany, 1993); and ({\it d}) since a more
negative slope at the high-mass end of the IMF would also increase the
predicted \ratzo\ and \co\ values, producing a conflict with the observations.

\par
On the other hand, observational evidence and arguments in favor of a change
at the low-mass end of the IMF, in the sense that the lower the metallicity
the larger the amount of small mass stars, have been presented previously
(Peimbert \& Serrano, 1982; Gusten \& Mezger, 1983; Larson, 1986; Richer et
al., 1991).
\par
In what follows we will consider that the $r$ values higher than one come from
varying the IMF at its low-mass end; for simplicity, we will assume that it is
the slope of the IMF which varies in the $0.01 < m <0.5$ range. In this case
$r$ can be computed as follows:
$$
r = {\int_{m_l}^{120}m\xi(m)dm \over \int_{m_l}^{120} m\xi'(m)dm},
\eqno (4.1)
$$
\noindent where $m_l$ is any value greater than 0.5, $\xi(m)$ is the KTG IMF,
and $\xi'(m)$ is a slope-modified IMF; both IMFs are normalized to one. Our
results produce a slope for $\xi'(m)$ in the $0.01 < m < 0.5$ interval, whose
absolute value increases with $r$: for the $r > 1$ values shown in the first
five rows of Table 2 we obtained, for a decreasing $r$, the following values
for the slope: $- 2.38$, $- 2.32$, $- 2.21$, $- 2.01$, and $- 1.76$. For the
model with
the S IMF and \mco\ $= 60$, an $r$ value of 3.20 is obtained, very similar to
the $r$ value derived with the KTG IMF. The value 0.5 for $m_l$ comes from one
of the points where the KTG IMF changes its slope; however, this point may be
moved up to 1 (adjusting the slope correspondingly) without any
change in the predicted abundance ratios.
\par
The change in \ox\ from $2.58 \times 10^{-3}$ to $1.62 \times 10^{-3}$ does
not affect significatively the O/\z, \ratzo, \dydzo, and \co\ ratios, the
largest change is for the \ratzo\ ratio and amounts to
3.6 \%. Thus, if an observational error in the determination of O exists and
is of the order of a factor of two, then the conclusions derived
from the abundance ratios will still be valid. In addition, the change in
the $r$ parameter is inversely proportional to the change in the O
abundance.
\par
Similarly, an analysis of the influence of the {\Large $\mu$} parameter on the
predictions by the models, shows that an increase in {\Large $\mu$} of 10 \%
produces a decrease in $r$ of 7.32 \%, and a decrease in {\Large $\mu$} of 10
\% produces an increase in $r$ of 7.96 \%. Furthermore, the changes in O/\z,
\ratzo, \dydzo, and \co\ are smaller than 0.5 \%; these models keep \mt\
constant and vary $M_{gas}$.
\par
\esp
\centerline{5. OUTFLOW MODELS}
\esp
\centerline{5.1 Outflow of well-mixed material}
\esp
The adopted log$ \mu$ and O values
can also be fitted by assuming that galaxies are ejecting
well-mixed material to the intergalactic medium via an ordinary
galactic wind. In chemical evolution models with the IRA assumption,
an ordinary wind is introduced by assuming an outflow modelled by
$f_O = \lambda (1-R)\psi$,
where $R$ is the mass fraction returned to the interstellar medium
by a generation of stars.
In our models, this wind is introduced as an outflow given by $f_O = \alpha
M_{gas}$, where the $\alpha$ parameter is then related to $\lambda$
through $\alpha= \lambda (1-R)\nu$. Notice that $R$ is ill
defined in the models that do not assume the IRA, nevertheless an
average $R$ value has been defined and used to compute the $\lambda $
parameter. Models with an outflow of well-mixed material, a KTG IMF, no black
holes, and $r$ = 1, are shown in Table 4. The first column shows
the different ages of the model, the second contains the ratio of the ejected
mass, $M_{gas}^e$, to the total mass, the third and fourth show
the values of the $\alpha $ and $\lambda $ parameters, the fifth shows the
mass fraction that remains trapped in stars, ($1 - R$), and the last four
columns show the
O/\z, \ratzo, $\Delta Y/\Delta \ox$, and \co\ values, respectively.
\par
The \dydzo\ values of the models in Table 4 are larger than those predicted
by closed-box models due to the delay of the He production relative to that
of O; this delay causes a preferential O loss to the intergalactic medium.
\par
The increase in the \dydzo\ ratios produced by outflow of well-mixed material
relative to those produced by closed-box models goes in the right direction
but is not large enough to explain the observational constraint; this is one
of the reasons that has led to the idea of a different type of galactic
outflow.

\esp
\centerline{5.2 Outflow of O-rich material}
\esp
It has been proposed that an outflow of O-rich material is present,
at least in some galaxies, to explain: ({\it a}) the small yield
in heavy elements seen in irregular galaxies
(Lequeux 1989; Tosi 1994; Peimbert et al. 1994a), ({\it b})
the large helium abundances
derived in some irregular and blue compact galaxies
(Aparicio, Garc\'ia-Pelayo, \& Moles 1988), ({\it c}) the large
\dydzo\ ratios derived from samples of galaxies
(Aparicio et al. 1988; Lequeux 1989; Pilyugin 1993; Tosi 1994;
Peimbert et al. 1994a; Marconi,  Matteucci \& Tosi, 1994), ({\it d}) the high
Fe/O ratio in the Magellanic Clouds
(Russell, Bessell, \& Dopita 1988a,b; Lequeux 1989)
and ({\it e}) the $Z$-Mass relation present
in irregular galaxies (De Young \& Gallagher 1990).
\par
By assuming that a fraction $\gamma$ of the mass of a SN
is ejected to the intergalactic medium without mixing with the
interstellar gas, several models with a KTG IMF for different values of the
parameter $\gamma$ have been calculated in order to fit the adopted
log$ \mu$ and O values, these models are shown in Tables 5 and 6
and Figures 4, 5, and 6, where the curves are non-linear fits to the computed
models (solid squares). The inclusion of SNe I
in these models produces a situation analogous to the one produced for the
closed models and the resulting predictions are similar to the ones indicated
by a dot in Figures 1, 2, and 3.
\par
For a galactic age of 1 Gyr, about half of the freshly made He is ejected by
stars in the 7.5--120 \mdot\
range and the other half, by stars in the 1.85--7.5 \mdot\ range, whereas
almost all of the \ox, is produced by high-mass stars; meanwhile for a galactic
age of 10 Gyr, stars in the 0.9--7.5 \mdot\ range produce about twice the He
produced by high-mass stars. Low mass stars end their lives as white dwarfs
and eject part of their masses in normal stellar winds during the red giant
phase and as superwinds during the preplanetary nebula phase; on the other
hand, stars in the 7.5--120 \mdot\ range, lose
part of their masses via stellar winds and afterwards explode
like SNe II, these ejecta have typical initial velocities
of a few thousands of \vel, and so a fraction of the ejected mass might be
expected to leave the parental galaxy.
\par
The first column of
Table 5 shows the galactic age; the chosen gamma values and the
corresponding $r$ values are shown in the second and third columns,
respectively; the fourth, fifth, sixth, and seventh columns show the
computed values of $\ox$/\z,  \dydzo, \ratzo, and \co, respectively; finally,
the last column shows the ratio of the ejected mass, $M_{gas}^e$, to the total
mass. As one would expect the \dydzo\ value increases
if the $\gamma$ value increases. The gamma value with $r = 1$ means that the
corresponding O-rich wind is sufficient to reproduce the adopted log$ \mu$
and O values; i. e., there is no need
for a modified IMF at the low-mass end. A value of $r$ smaller than one
indicates that the fraction of stars at the low-mass end of the IMF is smaller
than that given by the KTG IMF.
\par
As can be seen from Figures 4 and 5,
for a given IMF, the dependence of
\dydzo\ and \co\ with $t_g$ is like the one reported above
for the $\gamma$-dependence of \dydzo: if $t_g$ increases, more C and
He produced by the LMS will be ejected to the ISM and so the \dydzo\ and
\co\ values increase; this
behaviour is not followed by \ratzo\ when going from
1 to 10 Gyr (Fig. 6) since most of the (\z\ $-$ C $-$ O) elements are ejected
by stars with $m_i > 1.85$. If $\gamma $ increases, then $r$
decreases and the number of stars that participate in the enrichment of the
ISM increases, therefore, the production rate of any element increases. The
non-linear fits to the models in Figures 4, 5, and 6 were obtained assuming
that the logarithm of the ratio of any element to O varies as $a + {b/[(1 -
\gamma) + c]}$, where $a$, $b$, and $c$ are numbers to be determined for each
element.
\par
It follows from Figures 4, 5, and 6 that there is not a unique $\gamma $
value that fits the three observational constraints at a given age. As in the
case of the closed-box models, two possibilities can be followed: ({\it a}) to
adopt models that fit the \dydzo\ but not the \co\ and \ratzo\ constraints
(those with high $\gamma $ and low $r$ values, \eg, $\gamma = 0.6$ and $r$ =
1.4 for a $t_g$ = 10 Gyr), or ({\it b}) to adopt the models that reproduce de
\co\ and \ratzo\ but not the \dydzo\ constraints (those with low $\gamma $
and high $r$ values, \eg, $\gamma = 0$ and $r$ = 3.39 for the same $t_g$ =
10 Gyr). It should be mentioned, however, that there is a model ($\gamma
\sim 0.23$ and an $r \sim 2.66$ for a $t_g = 10$ Gyr) which barely fits the
three
observational constraints; note that this model requires a slope of $- 2.25$
for the IMF in the $0.01 < m_i < 0.5$ interval, a value which is steeper than
$- 1.3$, the
slope of the IMF for the solar vicinity in the same interval.
This result for metal-poor
galaxies supports the idea that the fraction of small-mass stars does
decrease with increasing metallicity (Peimbert \& Serrano, 1982; Gusten \&
Mezger, 1983; Larson, 1986; Richer et al., 1991).
\par
Table 6 is the analogous of Table 3 including two additional columns:
the second and the fifth, that show the $\gamma $ values and the galactic
wind contributions by gas mass, respectively.
\vfill\eject
\esp
\centerline{6. CONCLUSIONS}
\esp
We present chemical evolution models for irregular galaxies in order to try
to fit five observational constraints. We study the parameter space provided
by: ({\it a}) $t_g$, \ie, without IRA; ({\it b}) closed-box models varying
\mco, for each \mco\ a different $r$ is needed to match log$ \mu$ and O;
({\it c}) well-mixed outflow models with $r = 1$; and ({\it d}) O-rich
outflow models varying $\gamma $, for each $\gamma $ a different $r$ is needed
to match log$ \mu$ and O.
\par
As various authors have previously pointed out, closed models with a solar
vicinity IMF do not explain the low O abundances.
\par
The low O abundances require a change in the IMF: either a larger amount of
low-mass stars or a change in the slope at the high-mass end. We present
arguments against a change in the slope of the IMF at the high-mass end and
conclude that the change should be at the low-mass end according to the $r$
parameter defined by equation (4.1).
\par
The production of black holes that prevent stars with masses higher than \mbh\
from enriching the ISM can explain {\Large $\mu $}, O, and \dydzo\ with an
$r = 1.97$ and \mco $= 16.4$. This model fails to explain the \co\ and
\ratzo\ observational constraints.
We consider that black holes do not play an important role in the chemical
evolution of irregular galaxies due to their failure at explaining \co\ and
\ratzo, and also due to the results of chemical evolution models for the
solar neighborhood that do not require them.
\par
The best closed models are those without black holes, and among them, the one
with $t_g = 10$ Gyr, a KTG IMF, and $r = 3.39$ is best since: ({\it a}) there
is evidence of star formation during about 10 Gyr in Magellanic irregulars,
({\it b}) it explains {\Large $\mu $}, O, \co, and \ratzo, and ({\it c}) it
is the model that comes closer to explaining the \dydzo\ constraint.
\par
Errors or changes in O and {\Large $\mu $} of a factor of two or less,
produce changes smaller than 4 \% in the model-predicted \dydzo, \co, and
\ratzo\ values. Alternatively, the changes in $r$ are larger and: ({\it a})
are inversely proportional to the changes in O, and ({\it b}) proportional
to changes in $log(\mu^{-1})$.
\par
Our models have been obtained trying to fit the observational constraints of
a `typical' irregular galaxy, defined as the average of the properties of the
ten selected galaxies in Table 1. Nevertheless, our models are robust and
can be generalized to individual galaxies with changes of about a factor of
two in O and a factor of two in {\Large $\mu $}, relative to our `typical'
irregular. This range of values includes most of the galaxies in Table 1 with
the exception of the two metal poorest that might need a different set of
chemical evolution models and that were mainly used to determine $Y_p$.
\par
We computed outflow models with well-mixed material and $r = 1$ that fit
{\Large $\mu $}, O, \co, and \ratzo. These models predict \dydzo\ values closer
to the observed ones but not close enough; they also predict large $M_{gas}^e
/M_T$ values, the ejected gas however, has probably dissipated for a $t_g = 10$
Gyr model and would be difficult to detect. Well-mixed outflow models made
to fit {\Large $\mu $} and O with $r$ values between 1 and those of the closed
models would yield \dydzo, \co, and \ratzo\ values intermediate between those
presented in Table 4 and the closed-box models.
\par
Outflow models of O-rich material with $\gamma = 0.23$ and $r = 2.66$ can
marginally fit {\Large $\mu $}, O, \dydzo, \co, and \ratzo. Lower values of
$\gamma $ (requiring $r > 2.66$ values) fail to fit \dydzo; while larger values
of $\gamma $ (requiring $r < 2.66$ values) fail to fit \co\ and \ratzo.
A better agreement with the three abundance ratios could be obtained for a
smaller observed value of
\dydzo, and this could be so if the effect on the O abundance due to the
temperature structure of the H II regions is higher than the one adopted in
this study.
\par
The best fit to the five observational constraints is given by O-rich outflows
with $\gamma = 0.23$ and $r = 2.66$. For this model we obtain a slope of
$- 2.25$
for the low-mass end of the IMF; this slope is intermediate between the KTG
IMF slope of $- 1.3$ and a slope smaller than $- 2.5$ derived by Richer et
al. (1991) for globular clusters. The metallicity of our `typical' irregular
galaxy is intermediate between that of the solar vicinity and those of the
globular clusters, the change in the slope could therefore be due to a
metallicity effect.
\par
\esp
The authors express their gratitude to D. R. Garnett for
reading the manuscript and for his excellent suggestions, and to
J. S. Mathis for fruitful discussions. They also acknowledge a
thorough arbitration by B. E. J. Pagel, whose criticism and comments helped
to improve and clarify the manuscript.
\par
\vfill\eject
\centerline{REFERENCES}
\esp

\ref Aparicio, A., Garc{\'i}a-Pelayo, J. M., \& Moles, M. 1988, A\&AS, 74, 375

\ref Bi\'emont, E., Baudoux, M., Kurucz, R. L., Ansbacher, W., \& Pinnington,
E. H. 1991, A\&A, 249, 539

\ref Campbell, A. 1988, ApJ, 335, 644

\ref Carigi, L. 1994, ApJ, 424, 181

\ref De Young, D. S., \& Gallagher, J. S. 1990, ApJ, 356, L15

\ref de Vaucouleurs, G., de Vaucouleurs, A., Corwin, H. G., Buta, R. J.,
Paturel, G., \& Fouqu\'e, P. 1991, Third Reference Catalogue of Bright
Galaxies (New York:Springer-Verlag)

\ref Draine, B. T. \& Lee, H. M. 1984, ApJ, 285, 89

\ref Dufour, R. J., Shields, G. A., \& Talbot, R. J. 1981, ApJ, 252,
461

\ref Dufour, R. J., Walter, D. K., Garnett, D. R., Skillman, E. D., Peimbert,
M., Torres-Peimbert, S., Terlevich, R. J., Terlevich, E., Shields, G. A., \&
Mathis, J. S. 1994, in preparation

\ref Garnett, D. R. 1990, ApJ, 363, 142

\ref Garnett, D. R., Skillman, E. D., Dufour, R. J., Peimbert, M.,
Torres-Peimbert, S., Terlevich, R. J., Terlevich, E., \& Shields, G. A. 1994,
ApJ, submitted

\ref Gonz\'alez-Delgado, R. M., P\'erez, E., Tenorio-Tagle, G., V\'ilchez, J.
M., Terlevich, E., Terlevich, R., Telles, E., Rodr\'iguez-Espinosa, J. M.,
Mas-Hesse, M., Garc{\'ia}-Vargas, M. L., D{\'i}az, A. I., Cepa, J., \&
Casta\~neda, H. O. 1994, ApJ, in press

\ref Grevesse, N., \& Anders, E. 1989, in Cosmic Abundances of Matter, ed. J.
Waddington (New York: American Institute of Physics), 1

\ref Grevesse, N., Lambert, D. L., Sauval, A. J., van Dishoeck, E. F., Farmer,
C. B., \& Norton, R. H. 1990, A\&A, 232, 225

\ref Gusten, R., \& Mezger, P. G. 1983, Vistas Astr., 26, 159

\ref Hannaford, P., Lowe, R. M., Grevesse, N., \& Noels, A. 1992, A\&A, 259,
301

\ref Holweger, H., Bard, A., Kock, A., \& Kock, M. 1991, A\&A, 249, 545

\ref Hutchmeier, W. K., \& Richter, O. -G. 1988, A\&A, 203, 237

\ref Isotov, Y. I., Thuan, T. X., \& Lipovetsky, V. A. 1994, ApJ, in press

\ref Kroupa, P., Tout, C. A., \& Gilmore, G. 1993, MNRAS, 262, 545 (KTG)

\ref Kunth, K. D., \& Sargent, W. L. W. 1983, ApJ 273, 81

\ref Larson, R. B. 1986, MNRAS, 218, 409

\ref Larson, R. B., \& Tinsley, B. M. 1978, ApJ, 219, 46

\ref Lequeux, J. 1989, in Evolution of Galaxies-Astronomical
Observations, ed. I. Appenzeller, H. J. Habing, \& P. Lena (Springer), p. 147

\ref Lequeux, J., Peimbert, M., Rayo, J., Serrano, A., \&
Torres-Peimbert, S. 1979, A\&A, 80, 155

\ref Maeder, A. 1991, Pr\'e-Publications de L'Observatoire de Gen\`eve, S\'erie
C, 84

\ref --------------- 1992, A\&A, 264, 105

\ref --------------- 1993, A\&A 268, 833

\ref Marconi, G., Matteucci, F., \& Tosi, M. 1994, MNRAS, in press

\ref Massey, P., Garmany, C. D., Silkey, M., \& DeGioia-Eastwood, D.K. 1989a,
AJ, 97, 107

\ref Massey, P., Parker, J. Wm., \& Garmany, C. D. 1989b,
AJ, 98, 1305

\ref Mathis, J. S. 1994, ApJ, 422, 176

\ref Mathis, J. S., Rumpl, W. \& Nirdsieck, K. H. 1977, ApJ, 217, 425

\ref McGaugh, S. S. 1991, ApJ, 380, 140

\ref Nomoto, K., Shigeyama, T., \& Tsujimoto, T. 1990, in IAU Symp. 145,
Evolution of Stars: The Photospheric Abundance Connection, ed. G. Michaud \& A.
Tutukov (San Francisco: Book Crafters), 21

\ref Pagel, B. E. J., 1987, in A Unified View of the Macro- \& the
Micro-cosmos, ed. A. de R\'ujula, D. V. Nanopoulos, and P. A. Shaver,
(Singapore: World Scientific), p. 399

\ref Pagel, B. E. J., Simonson, E. A., Terlevich, R. J., \& Edmunds, M. G.
1992, MNRAS, 255, 325 (PSTE)

\ref Parker, J. Wm., \& Garmany, C. D. 1993, AJ, 106, 1471

\ref Parker, J. Wm., Garmany, C. D., Massey, P., \& Walborn, N. R. 1992,
AJ, 103, 1205

\ref Peimbert, M. 1993, Rev. Mexicana Astron. Af., 27, 9

\ref Peimbert, M., Col{\'i}n, P., \& Sarmiento, A. 1994a, in Violent
Star Formation from 30-Doradus to QSOs, ed. G. Tenorio-Tagle (Cambridge
University Press), in press

\ref Peimbert, M., Pe\~na, M., \& Torres-Peimbert, S. 1986, A\&A, 158,
266

\ref Peimbert, M., Sarmiento, A., \& Col{\'i}n, P. 1994b, Rev. Mexicana
Astron. Af., 28, 181

\ref Peimbert, M., Sarmiento, A., \& Fierro, J. 1991, PASP, 103, 815

\ref Peimbert, M., \& Serrano, A. 1982, MNRAS, 198, 563

\ref Peimbert, M., Storey, P. J., \& Torres-Peimbert, S. 1993a, ApJ,
414, 626

\ref Peimbert, M., \& Torres-Peimbert, S, 1976, ApJ, 203, 581

\ref Peimbert M., Torres-Peimbert, S., \& Dufour, R. J. 1993b, ApJ,
418, 760

\ref Peimbert, M., Torres-Peimbert, S., \& Ruiz, M. T. 1992, Rev. Mexicana
Astron. Af., 24, 155

\ref Pilyugin, L. S. 1993, A\&A, 277, 42

\ref Prantzos, N., Vangioni-Flam, E. \& Chauveau, S. 1994, A\&A, 285, 132

\ref Renzini A., \& Voli, M. 1981, A\&A, 94, 175

\ref Richer, H. B., Fahlman, G. G., Buonanno, R., Fusi Pecci, F., Searle, L.,
\& Thompson, I. B. 1991, ApJ, 381, 147

\ref Rosa, M. R., \& Benvenuti, P. 1994, A\&A, in press

\ref Russell, S. C., Bessell, M. S., \& Dopita, M. A. 1988a, in Galactic  and
Extragalactic Star Formation, ed. R. E. Pudritz \& M. Fich (Kluwer), p. 106

\ref Russell, S. C., Bessell, M. S., \& Dopita, M. A. 1988b, in The Impact  of
Very High S/N Spectroscopy, ed. G. Cayrel de Strobel \& M. Spite (Kluwer), p.
545

\ref Salpeter, E. E. 1955, ApJ, 121, 161 (S)

\ref Scalo, J. M. 1986, Fund. Cosmic Phys., 11, 1

\ref Schaller, G., Schaerer, D., Meynet, G., \& Maeder, A. 1992, A\&ASS, 96,
269

\ref Schmidt, K. H., \& Boller, T. 1993, Astron. Nachr., 314, 361

\ref Skillman, E. D. 1991, PASP, 103, 919

\ref Skillman, E. D., \& Kennicutt, R. C. 1993, ApJ, 411, 655

\ref Skillman, E. D., Kennicutt, R. C., \& Hodge, P. W. 1989, ApJ, 347,
875

\ref Skillman, E. D., Terlevich, R. J., Kennicut, R. C., Garnett, D. R.,
\& Terlevich, E. 1994, ApJ, 431, 172

\ref Staveley-Smith, L., Davies, R. D., \& Kinman, T. D. 1992, MNRAS,
258, 334

\ref Torres-Peimbert, S., Peimbert, M., \& Fierro, J. 1989, ApJ, 345,
 186

\ref Tosi, M. 1994, preprint

\ref Vigroux, L., Stasinska, G., \& Comte, G. 1987, A\&A, 172, 15
\par
\vfill\eject
\par
\null\esp
\centerline{FIGURE CAPTIONS}
\par
\esp
\esp
FIGURE 1. \dydzo\ {\it versus} $m_{\rm co}$ predicted by the closed models.
The $m_{\rm co}$ values that correspond to a given initial mass $m$ (or to a
given \mbh), are from Maeder(1992). The region allowed by
observations is shown by a shaded band. The point represents
a model with a KTG IMF, a galactic time scale of 10 Gyr, and including
SNe I of binary origin.
\par
\esp
FIGURE 2. \co\ {\it versus} $m_{\rm co}$ predicted by the
closed models. The shaded
band and the point have the same meaning as in Figure 1, \ie, the region
allowed by observations and the prediction when SNe I are included; different
lines follow also the convention of Figure 1.
\par
\esp
FIGURE 3. \ratzo\ {\it versus} $m_{\rm co}$ predicted by
the closed models. Regions, symbols, and notation have the same meaning
as in the two previous figures.
\par
\esp
FIGURE 4. $\Delta Y/\Delta {\rm O}$ {\it versus} $\gamma$ assuming
three different galactic ages. The models (solid squares) are adjusted by
non-linear fits given by $a + b/[(1-\gamma) + c]$ (solid lines). The shaded
band indicates the region allowed by observations.
\par
\esp
FIGURE 5.- Analogue of Figure 4 for \co.
\par
\esp
FIGURE 6. Analogue of Figure 4 for \ratzo. The 1 and 10 Gyr models are almost
identical in this case; consequently, only the 10 Gyr model is presented.
\par
\vfill\eject
\par
\null\vskip2cm
\hsize 6.3 in
\centerline {TABLE 1}
\vskip0.2 cm
\centerline{PROPERTIES OF SELECTED GALAXIES}
\vskip 0.5 cm
\hrule width 6 in
\vskip0.05 cm
\hrule width 6 in
\vskip0.2 cm
\settabs 7 \columns
\+ \ $Galaxy$ & \hfil $\log(M_T$/\mdot) & \hfil $\log(M_{gas}$/\mdot)
& \hfil $\log \mu $ & \hfil $Y$ & \hfil $10^3$O & \quad $10^3Z$  \cr
\vskip0.2cm
\hrule width 6 in
\vskip 0.2 cm
\+ I Zw 18 & \hfil \ 8.26 & \hfil 8.16 & \hfil -0.1: & \hfil 0.230 & \hfil
0.200 & \quad 0.370  \cr
\+ UGC 4483 & \hfil \ 7.94 & \hfil 7.84 & \hfil -0.1: & \hfil 0.239 & \hfil
0.403 & \quad 0.746  \cr
\+ Mrk 600 & \hfil \ 8.85 & \hfil 8.67 & \hfil -0.18 & \hfil 0.240 & \hfil
1.241 & \quad 2.297  \cr
\+ SMC & \hfil \ 9.18 & \hfil 8.80 & \hfil -0.38 & \hfil 0.237 & \hfil 1.431 &
\quad 2.647  \cr
\+ II Zw 40 & \hfil \ 8.87 & \hfil 8.40 & \hfil -0.47 & \hfil 0.251 & \hfil
1.686 & \quad 3.122 \cr
\+ IC 10 & \hfil \ 9.73 & \hfil 9.11 & \hfil -0.62 & \hfil 0.240 & \hfil 1.751
& \quad 3.242  \cr
\+ NGC 6822 & \hfil \ 9.23 & \hfil 8.28 & \hfil -0.95 & \hfil 0.246 & \hfil
2.085 & \quad 3.861  \cr
\+ II Zw 70 & \hfil \ 9.11 & \hfil 8.58 & \hfil -0.53 & \hfil 0.250 & \hfil
2.074 & \quad 3.841  \cr
\+ LMC & \hfil \ 9.78 & \hfil 8.85 & \hfil -0.93 & \hfil 0.250 & \hfil 2.608 &
\quad 4.829  \cr
\+ NGC 4449 & \hfil 10.60 & \hfil 9.78 & \hfil -0.82 & \hfil 0.251 & \hfil
2.856 & \quad 5.288  \cr
\vskip0.3cm
\+ Average$^a$ & \hfil \ 9.15 & \hfil 8.65 & \hfil -0.51 & \hfil 0.243 & \hfil
1.633 & \quad 3.023  \cr
\vskip 0.2 cm
\hrule width 6 in
\baselineskip = 6 pt
\parskip = -.3 pt
\hsize 6 in
\vskip 0.2cm
\par
\noindent{\ninerm $^a ${ \tenrm Linear Regressions: $Y = (0.232 \pm 0.004) +
(7.10 \pm 1.62) \times {\rm O}$, $Y = (0.232 \pm 0.004) + (3.83 \pm 0.87)
\times Z$}}
\par
\vfill\eject
\magnification 1200
\baselineskip=16pt
\par
\null\vskip 5 cm
\centerline {TABLE 2}
\vskip 0.3 cm
\centerline{CLOSED BOX MODELS$^a$}
\vskip 0.3 cm
\hrule width  5.7 in
\vskip 0.1 cm
\hrule width  5.7 in
\vskip 0.3 cm
\settabs 10 \columns
\+ \mco & \quad \mbh & \qquad \qquad $r$ & \quad \qquad He &
\qquad $10^3$C & \qquad $10^3$\z & \ \ \qquad $\rm O \over Z$ & \quad \ ${Z -
{\rm C} - \ox}\over \ox$ & \ \qquad $\Delta Y\over {\Delta \ox}$ & \ \ \qquad
${\rm C}\over \ox$\cr
\vskip 0.3 cm
\hrule width  5.7 in
\vskip 0.2 cm
\+ \quad $60$ & \quad no BH & \quad \qquad $3.39$ & \qquad $0.2376$ & \qquad
$0.619$ & \qquad $5.00$ & \qquad $0.516$ & \qquad $0.700$ & \qquad $2.946$ &
\qquad $0.240$ \cr
\+ \quad $40$ & $88.2-106$ & \quad \qquad $3.05$ & \qquad $0.2385$ & \qquad
$0.686$ & \qquad $5.28$ & \qquad $0.489$ & \qquad $0.780$ & \qquad $3.292$ &
\qquad $0.266$ \cr
\+ \quad $25$ & $60.4-74.8$ & \quad \qquad $2.47$ & \qquad $0.2406$ & \qquad
$0.848$ & \qquad $5.96$ & \qquad $0.433$ & \qquad $0.983$ & \qquad $4.108$ &
\qquad $0.329$ \cr
\+ \quad $15$ & $42.0-52.8$ & \quad \qquad $1.84$ & \qquad $0.2444$ & \qquad
$1.147$ & \qquad $7.28$ & \qquad $0.354$ & \qquad $1.377$ & \qquad $5.581$ &
\qquad $0.445$ \cr
\+ \quad $10$ & $31.6-41.8$ & \quad \qquad $1.38$ & \qquad $0.2497$ & \qquad
$1.578$ & \qquad $9.12$ & \qquad $0.283$ & \qquad $1.924$ & \qquad $7.636$ &
\qquad $0.612$ \cr
\+ \quad\ $7$ & $25.1-24.8$ & \quad \qquad $0.92$ & \qquad $0.2623$ & \qquad
$2.744$ & \qquad $13.4$ & \qquad $0.193$ & \qquad $3.111$ & \qquad $12.52$ &
\qquad $1.064$ \cr
\+ \quad\ $5$ & $20.1-20.1$ & \quad \qquad $0.69$ & \qquad $0.2774$ & \qquad
$4.572$ & \qquad $17.8$ & \qquad $0.144$ & \qquad $4.146$ & \qquad $18.37$ &
\qquad $1.772$ \cr
\vskip 0.1cm
\+ \quad $60$ & \quad no BH & \quad \qquad $5.29$ & \qquad $0.2348$ & \qquad
$0.379$ & \qquad $3.17$ & \qquad $0.510$ & \qquad $0.725$ & \qquad $2.970$ &
\qquad $0.234$ \cr
\vskip 0.2 cm
\hrule width  5.7 in
\vskip 0.3 cm
\baselineskip = 6 pt
\parskip = -.3 pt
\hsize 6 in
\noindent $^a ${ \tenrm For \ox\ $ = 2.580 \times 10^{-3}$, with the exception
of the last model which is for \ox\ $ = 1.616 \times 10^{-3}$}
\vfill\eject
\magnification 1200
\baselineskip=16pt
\par
\null\vskip 2 cm
\centerline{TABLE 3}
\bigskip
\centerline{ISM ABUNDANCES BUDGET FOR CLOSED MODELS$^a$}
\bigskip
\centerline{\vbox{ \tabskip=1pt {\halign{
& \hfil\quad{#}\hfil & \hfil\quad{#}\hfil & \hfil\quad{#}\hfil &
\hfil\quad{#}\hfil & \hfil\quad{#}\hfil &  \hfil\quad{#}\hfil   \cr
\tablerule\cr
\noalign{\vskip 1pt}
\tablerule\cr
\tablerule\cr
\noalign{\vskip 3pt}
$t_g$ & $i$ & SF & LMS & HMS$_{\rm W}$ & HMS$_{\rm {SN}}$ & $X_i$ \cr
(Gyr)& & & & & & \cr
\noalign{\vskip 3pt}
\tablerule\cr
\tablerule\cr
0.10 & He & -5.623$^.10^{-1}$ &  3.728$^.10^{-3}$ &   1.395$^.10^{-3}$ &
1.638$^.10^{-2}$ &  2.346$^.10^{-1}$ \cr
   & C  & -2.804$^.10^{-4}$ &  8.323$^.10^{-6}$ &   8.607$^.10^{-6}$ &
6.127$^.10^{-4}$ &  3.492$^.10^{-4}$ \cr
   & O  & -2.323$^.10^{-3}$ &  3.819$^.10^{-6}$ &   6.091$^.10^{-6}$ &
4.893$^.10^{-3}$ &  2.580$^.10^{-3}$ \cr
   & $Z$  & -3.835$^.10^{-3}$ &  2.331$^.10^{-4}$ &   2.062$^.10^{-4}$ &
7.877$^.10^{-3}$ &  4.481$^.10^{-3}$ \cr
\noalign{\vskip 4pt}
1.00 & He & -5.797$^.10^{-1}$ &  2.148$^.10^{-2}$ &   1.650$^.10^{-3}$ &
1.794$^.10^{-2}$ &  2.367$^.10^{-1}$ \cr
   & C  & -4.807$^.10^{-4}$ &  3.822$^.10^{-4}$ &   1.544$^.10^{-5}$ &
6.629$^.10^{-4}$ &  5.798$^.10^{-4}$ \cr
   & O  & -2.583$^.10^{-3}$ &  6.550$^.10^{-5}$ &   8.859$^.10^{-6}$ &
5.089$^.10^{-3}$ &  2.580$^.10^{-3}$ \cr
   & $Z$  & -4.867$^.10^{-3}$ &  1.348$^.10^{-3}$ &   2.991$^.10^{-4}$ &
8.192$^.10^{-3}$ &  4.972$^.10^{-3}$ \cr
\noalign{\vskip 4pt}
10.0 & He & -6.003$^.10^{-1}$ &  4.289$^.10^{-2}$ &   1.704$^.10^{-3}$ &
1.810$^.10^{-2}$ &  2.376$^.10^{-1}$ \cr
   & C  & -6.159$^.10^{-4}$ &  5.473$^.10^{-4}$ &   1.688$^.10^{-5}$ &
6.702$^.10^{-4}$ &  6.185$^.10^{-4}$ \cr
   & O  & -2.693$^.10^{-3}$ &  1.502$^.10^{-4}$ &   9.339$^.10^{-6}$ &
5.114$^.10^{-3}$ &  2.580$^.10^{-3}$ \cr
   & $Z$  & -5.256$^.10^{-3}$ &  1.725$^.10^{-3}$ &   3.191$^.10^{-4}$ &
8.217$^.10^{-3}$ &  5.004$^.10^{-3}$ \cr
\noalign{\vskip 5pt}
10.0 & He & -5.793$^.10^{-1}$ & 2.654$^.10^{-2}$ & 8.107$^.10^{-4}$ &
1.142$^.10^{-2}$ &
 2.348$^.10^{-1}$ \cr
     &  C & -3.690$^.10^{-4}$ & 3.311$^.10^{-4}$ & 2.622$^.10^{-6}$ &
4.145$^.10^{-4}$ &
 3.792$^.10^{-4}$ \cr
     &  O & -1.625$^.10^{-3}$ & 6.051$^.10^{-5}$ & 2.254$^.10^{-6}$ &
3.178$^.10^{-3}$ &
 1.616$^.10^{-3}$ \cr
     &$Z$ & -3.194$^.10^{-3}$ & 1.006$^.10^{-3}$ & 8.889$^.10^{-5}$ &
5.266$^.10^{-3}$ &
 3.167$^.10^{-3}$ \cr
\noalign{\vskip 3pt}
\tablerule\cr
\tablerule\cr
}}}}
\hskip0.05cm{$^a ${ \tenrm Adjusting O $ = 2.580 \times 10^{-3}$, with the
exception of the last model which adjusts}
\par
\hskip0.2cm{ \tenrm O $ = 1.616 \times 10^{-3}$}
\par
\vfill\eject
\par
\null\vskip 5 cm
\centerline{TABLE 4}
\medskip
\centerline{OUTFLOW  OF WELL-MIXED MATERIAL$^a$}
\bigskip
\centerline{\vbox{ \tabskip=1pt {\halign{
& \hfil\quad{#}\hfil & \hfil\quad{#}\hfil & \hfil\quad{#}\hfil &
\hfil\quad{#}\hfil & \hfil\quad{#}\hfil &  \hfil\quad{#}\hfil   \cr
\tablerule\cr
\noalign{\vskip 1pt}
\tablerule\cr
\tablerule\cr
\noalign{\vskip 4pt}
$t_g$ & $M_{gas}^e \over M_T$ & $\alpha$ & $\lambda$ & $(1-R)$ & $\rm O\over Z$
& ${Z-\rm C-\rm O}\over \rm O$ & $\Delta Y\over \Delta \rm O$ & $\rm C\over
\rm O$ \cr
\noalign{\vskip 1pt}
(Gyr) & & (Gyr$^{-1}$) & & & & & & \cr
\noalign{\vskip 4pt}
\tablerule\cr
\tablerule\cr
0.10 & 6.43 & 29.40 & 10.77 & 0.798 & 0.505 & 0.821 & 2.861 & 0.159  \cr
1.00 & 6.50 & 2.932 & 10.13 & 0.742 & 0.486 & 0.746 & 3.290 & 0.312  \cr
10.0 & 7.95 & 0.316 & 13.13 & 0.606 & 0.512 & 0.703 & 3.347 & 0.250 \cr
\noalign{\vskip 1pt}
10.0 & 16.07 & 0.390 & 27.24 & 0.587 & 0.506 & 0.729 & 3.530 & 0.247 \cr
\noalign{\vskip 4pt}
\tablerule\cr
\tablerule\cr
}}}}
\noindent $^a$ For O$ = 2.580 \times 10^{-3}$ with the exception of the last
model which is for O$ = 1.616 \times 10^{-3}$
\par
\vfill\eject
\par
\null\vskip 2 cm
\centerline{TABLE 5}
\medskip
\centerline{OUTFLOW OF O-RICH MATERIAL$^a$}
\bigskip
{\medtype
\centerline{\vbox{ \tabskip=1pt {\halign{
& \hfil\quad{#}\hfil & \hfil\quad{#}\hfil & \hfil\quad{#}\hfil &
\hfil\quad{#}\hfil & \hfil\quad{#}\hfil &  \hfil\quad{#}\hfil   \cr
\tablerule\cr
\noalign{\vskip 1pt}
\tablerule\cr
\tablerule\cr
\noalign{\vskip 1pt}
$t_g$ & $\gamma$ & $r$ & $\rm O \over Z$ & $\Delta Y\over \Delta \rm O$ &
${Z-\rm C-\rm O}\over \rm O$ & $\car \over \ox$ &
$M_{gas}^e/M_T$ \cr
\noalign{\vskip 1pt}
(Gyr) & & & & & & & ($10^{-2}$)\cr
\noalign{\vskip 4pt}
\tablerule\cr
\tablerule\cr
0.10 & 0.00 & 3.240 & 0.576 & 1.783 & 0.601 & 0.135 & 0.910 \cr
     & 0.25 & 2.442 & 0.562 & 1.938 & 0.642 & 0.137 & 1.366 \cr
     & 0.50 & 1.644 & 0.536 & 2.287 & 0.723 & 0.141 & 2.276 \cr
     & 0.75 & 0.847 & 0.467 & 3.372 & 0.988 & 0.155 & 5.159 \cr
     & 0.90 & 0.370 & 0.315 & 7.016 & 1.939 & 0.231 & 14.34 \cr
\noalign{\vskip 4pt}
1.00 & 0.00 & 3.290 & 0.519 & 2.597 & 0.702 & 0.225 & 0.910 \cr
     & 0.25 & 2.497 & 0.489 & 3.023 & 0.787 & 0.257 & 1.442 \cr
     & 0.50 & 1.704 & 0.437 & 3.953 & 0.963 & 0.325 & 2.428 \cr
     & 0.75 & 0.910 & 0.321 & 6.860 & 1.566 & 0.553 & 5.539 \cr
     & 0.90 & 0.453 & 0.165 &15.659 & 3.470 & 1.604 & 14.72 \cr
\noalign{\vskip 4pt}
10.0 & 0.00 & 3.390 & 0.516 & 2.946 & 0.700 & 0.240 & 0.910 \cr
     & 0.25 & 2.599 & 0.485 & 3.527 & 0.784 & 0.277 & 1.442 \cr
     & 0.50 & 1.807 & 0.431 & 4.729 & 0.966 & 0.356 & 2.656 \cr
     & 0.75 & 1.013 & 0.331 & 8.411 & 1.593 & 0.622 & 5.615 \cr
     & 0.90 & 0.562 & 0.157 &19.380 & 3.519 & 1.853 & 14.64 \cr
\noalign{\vskip 6pt}
10.0 & 0.00 & 5.291 & 0.510 & 2.970 & 0.725 & 0.235 & 0.910 \cr
     & 0.25 & 4.026 & 0.484 & 3.527 & 0.798 & 0.270 & 1.214 \cr
     & 0.50 & 2.760 & 0.437 & 4.641 & 0.948 & 0.341 & 1.897 \cr
     & 0.75 & 1.495 & 0.331 & 8.045 & 1.456 & 0.561 & 3.794 \cr
     & 0.90 & 0.744 & 0.175 &18.750 & 3.376 & 1.351 & 9.712 \cr
\noalign{\vskip 4pt}
\tablerule\cr
\tablerule\cr
}}}}
}
\hskip0.2cm{$^a $ For O $ = 2.580 \times 10^{-3}$ with the exception of the
last five models which are}
\par
\hskip0.5cm{for O $ = 1.616 \times 10^{-3}$}
\par
\vfill\eject
\centerline{TABLE 6}
\vskip 1pt
\centerline{ISM ABUNDANCES BUDGET}
\vskip 1pt
\centerline{MODELS WITH AN OUTFLOW OF O-RICH MATERIAL}
\vskip 2pt
{\medtype \centerline{\vbox{ \tabskip=1pt {\halign{
& \hfil\quad{#}\hfil & \hfil\quad{#}\hfil & \hfil\quad{#}\hfil &
\hfil\quad{#}\hfil & \hfil\quad{#}\hfil &  \hfil\quad{#}\hfil   \cr
\tablerule\cr
\noalign{\vskip 1pt}
\tablerule\cr
\tablerule\cr
\noalign{\vskip 1pt}
$t_g$ & $\gamma$ & $i$ & SF & GW & LMS & HMS$_{\rm W}$ & HMS$_{\rm SN}$ & $X_i$
\cr
(Gyr) & & & & & & & & \cr
\noalign{\vskip 2pt}
\tablerule\cr
\tablerule\cr
0.10 & 0.25 & He & -5.670$^. 10^{-1}$ & -5.466$^. 10^{-3}$ & 4.992$^. 10^{-3}$
& 1.876$^. 10^{-3}$ & 2.190$^. 10^{-2}$ & 2.350$^. 10^{-1}$ \cr
     &      & C  & -2.839$^. 10^{-4}$ & -2.048$^. 10^{-4}$ & 1.114$^. 10^{-5}$
& 1.195$^. 10^{-5}$ & 8.195$^. 10^{-4}$ & 3.538$^. 10^{-4}$ \cr
     &      & O  & -2.342$^. 10^{-3}$ & -1.636$^. 10^{-3}$ & 5.118$^. 10^{-6}$
& 8.281$^. 10^{-6}$ & 6.544$^. 10^{-3}$ & 2.580$^. 10^{-3}$ \cr
     &      & $Z$  & -3.901$^. 10^{-3}$ & -2.632$^. 10^{-3}$ & 3.121$^.
10^{-4}$ & 2.798$^. 10^{-4}$ & 1.053$^. 10^{-2}$ & 4.589$^. 10^{-3}$ \cr
\noalign{\vskip 2pt}
     & 0.50 & He & -5.767$^. 10^{-1}$ & -1.650$^. 10^{-2}$ & 7.555$^. 10^{-3}$
& 2.861$^. 10^{-3}$ & 3.307$^. 10^{-2}$ & 2.359$^. 10^{-1}$ \cr
     &      & C  & -2.911$^. 10^{-4}$ & -6.184$^. 10^{-4}$ & 1.686$^. 10^{-5}$
& 1.942$^. 10^{-5}$ & 1.237$^. 10^{-3}$ & 3.638$^. 10^{-4}$ \cr
     &      & O  & -2.381$^. 10^{-3}$ & -4.940$^. 10^{-3}$ & 7.760$^. 10^{-6}$
& 1.293$^. 10^{-5}$ & 9.880$^. 10^{-3}$ & 2.580$^. 10^{-3}$ \cr
     &      & $Z$  & -4.038$^. 10^{-3}$ & -7.940$^. 10^{-3}$ & 4.724$^.
10^{-4}$ & 4.349$^. 10^{-4}$ & 1.588$^. 10^{-2}$ & 4.810$^. 10^{-3}$ \cr
\noalign{\vskip 2pt}
     & 0.75 & He & -6.075$^. 10^{-1}$ & -5.052$^. 10^{-2}$ & 1.554$^. 10^{-2}$
& 6.024$^. 10^{-3}$ & 6.749$^. 10^{-2}$ & 2.387$^. 10^{-1}$ \cr
     &      & C  & -3.155$^. 10^{-4}$ & -1.893$^. 10^{-3}$ & 3.467$^. 10^{-5}$
& 4.949$^. 10^{-5}$ & 2.525$^. 10^{-3}$ & 4.002$^. 10^{-4}$ \cr
     &      & O  & -2.506$^. 10^{-3}$ & -1.512$^. 10^{-2}$ & 1.605$^. 10^{-5}$
& 2.937$^. 10^{-5}$ & 2.016$^. 10^{-2}$ & 2.580$^. 10^{-3}$ \cr
     &      & $Z$  & -4.490$^. 10^{-3}$ & -2.423$^. 10^{-2}$ & 9.719$^.
10^{-4}$ & 9.676$^. 10^{-4}$ & 3.231$^. 10^{-2}$ & 5.530$^. 10^{-3}$ \cr
\noalign{\vskip 4pt}
10.0 & 0.25 & He & -6.181$^. 10^{-1}$ & -6.035$^. 10^{-3}$ & 5.759$^. 10^{-2}$
& 2.368$^. 10^{-3}$ & 2.422$^. 10^{-2}$ & 2.391$^. 10^{-1}$ \cr
     &      & C  & -7.262$^. 10^{-4}$ & -2.250$^. 10^{-4}$ & 7.403$^. 10^{-4}$
& 2.652$^. 10^{-5}$ & 9.001$^. 10^{-4}$ & 7.158$^. 10^{-4}$ \cr
     &      & O  & -2.771$^. 10^{-3}$ & -1.712$^. 10^{-3}$ & 2.013$^. 10^{-4}$
& 1.363$^. 10^{-5}$ & 6.847$^. 10^{-3}$ & 2.580$^. 10^{-3}$ \cr
     &      & $Z$  & -5.695$^. 10^{-3}$ & -2.739$^. 10^{-3}$ & 2.331$^.
10^{-3}$ & 4.631$^. 10^{-4}$ & 1.096$^. 10^{-2}$ & 5.318$^. 10^{-3}$ \cr
\noalign{\vskip 2pt}
     & 0.50 & He & -6.564$^. 10^{-1}$ & -1.829$^. 10^{-2}$ & 8.791$^. 10^{-2}$
& 3.871$^. 10^{-3}$ & 3.669$^. 10^{-2}$ & 2.422$^. 10^{-1}$ \cr
     &      & C  & -9.699$^. 10^{-4}$ & -6.867$^. 10^{-4}$ & 1.147$^. 10^{-3}$
& 5.444$^. 10^{-5}$ & 1.374$^. 10^{-3}$ & 9.187$^. 10^{-4}$ \cr
     &      & O  & -2.943$^. 10^{-3}$ & -5.193$^. 10^{-3}$ & 3.063$^. 10^{-4}$
& 2.470$^. 10^{-5}$ & 1.039$^. 10^{-2}$ & 2.580$^. 10^{-3}$ \cr
     &      & $Z$  & -6.679$^. 10^{-3}$ & -8.245$^. 10^{-3}$ & 3.605$^.
10^{-3}$ & 8.172$^. 10^{-4}$ & 1.649$^. 10^{-2}$ & 5.991$^. 10^{-3}$ \cr
\noalign{\vskip 2pt}
     & 0.75 & He & -7.718$^. 10^{-1}$ & -5.609$^. 10^{-2}$ & 1.841$^. 10^{-1}$
& 9.868$^. 10^{-3}$ & 7.502$^. 10^{-2}$ & 2.517$^. 10^{-1}$ \cr
     &      & C  & -1.874$^. 10^{-3}$ & -2.154$^. 10^{-3}$ & 2.517$^. 10^{-3}$
& 2.426$^. 10^{-4}$ & 2.873$^. 10^{-3}$ & 1.604$^. 10^{-3}$ \cr
     &      & O  & -3.458$^. 10^{-3}$ & -1.595$^. 10^{-2}$ & 6.321$^. 10^{-4}$
& 8.602$^. 10^{-5}$ & 2.127$^. 10^{-2}$ & 2.580$^. 10^{-3}$ \cr
     &      & $Z$  & -1.028$^. 10^{-2}$ & -2.476$^. 10^{-2}$ & 7.874$^.
10^{-3}$ & 2.447$^. 10^{-3}$ & 3.301$^. 10^{-2}$ & 8.294$^. 10^{-3}$ \cr
\noalign{\vskip 5pt}
10.0 & 0.25 & He &
-5.903$^. 10^{-1}$ &  -3.805$^. 10^{-3}$ &   3.553$^. 10^{-2}$ &   1.112$^.
10^{-3}$ &   1.527$^. 10^{-2}$ &
 2.357$^. 10^{-1}$ \cr
     &      & C  &
-4.295$^. 10^{-4}$ &  -1.387$^. 10^{-4}$ &   4.452$^. 10^{-4}$ &   4.184$^.
10^{-6}$ &   5.550$^. 10^{-4}$ &
 4.362$^. 10^{-4}$ \cr
     &      & O  &
-1.655$^. 10^{-3}$ &  -1.062$^. 10^{-3}$ &   8.102$^. 10^{-5}$ &   3.208$^.
10^{-6}$ &   4.249$^. 10^{-3}$ &
 1.616$^. 10^{-3}$ \cr
     &      & $Z$  &
-3.411$^. 10^{-3}$ &  -1.756$^. 10^{-3}$ &   1.353$^. 10^{-3}$ &   1.309$^.
10^{-4}$ &   7.023$^. 10^{-3}$ &
 3.341$^. 10^{-3}$ \cr
\noalign{\vskip 2pt}
     & 0.50 & He &
-6.121$^. 10^{-1}$ &  -1.146$^. 10^{-2}$ &   5.371$^. 10^{-2}$ &   1.765$^.
10^{-3}$ &   2.300$^. 10^{-2}$ &
 2.375$^. 10^{-1}$ \cr
     &      & C  &
-5.566$^. 10^{-4}$ &  -4.197$^. 10^{-4}$ &   6.790$^. 10^{-4}$ &   8.682$^.
10^{-6}$ &   8.397$^. 10^{-4}$ &
 5.510$^. 10^{-4}$ \cr
     &      & O  &
-1.715$^. 10^{-3}$ &  -3.203$^. 10^{-3}$ &   1.226$^. 10^{-4}$ &   5.508$^.
10^{-6}$ &   6.406$^. 10^{-3}$ &
 1.616$^. 10^{-3}$ \cr
     &      & $Z$  &
-3.868$^. 10^{-3}$ &  -5.269$^. 10^{-3}$ &   2.063$^. 10^{-3}$ &   2.351$^.
10^{-4}$ &   1.054$^. 10^{-2}$ &
 3.699$^. 10^{-3}$ \cr
\noalign{\vskip 2pt}
    & 0.75 & He &
-6.801$^. 10^{-1}$ &  -3.479$^. 10^{-2}$ &   1.100$^. 10^{-1}$ &   4.171$^.
10^{-3}$ &   4.653$^. 10^{-2}$ &
 2.430$^. 10^{-1}$ \cr
     &      & C &
-9.903$^. 10^{-4}$ &  -1.291$^. 10^{-3}$ &   1.428$^. 10^{-3}$ &   3.823$^.
10^{-5}$ &   1.722$^. 10^{-3}$ &
 9.068$^. 10^{-4}$ \cr
     &      & O &
-1.901$^. 10^{-3}$ &  -9.748$^. 10^{-3}$ &   2.505$^. 10^{-4}$ &   1.675$^.
10^{-5}$ &   1.300$^. 10^{-2}$ &
 1.616$^. 10^{-3}$ \cr
     &      & $Z$ &
-5.453$^. 10^{-3}$ &  -1.580$^. 10^{-2}$ &   4.335$^. 10^{-3}$ &   7.251$^.
10^{-4}$ &   2.106$^. 10^{-2}$ &
 4.876$^. 10^{-3}$ \cr
\noalign{\vskip 2pt}
\tablerule\cr
\tablerule\cr
}}}}}
\end